\documentclass[preprint,12pt]{elsarticle}
\usepackage{amsmath}
\usepackage{hyperref} 

\newcommand{\euler}{\operatorname{e}}
\newcommand{\ee}[1]{{\euler}^{#1}}

\usepackage{bm}

\usepackage{graphicx}
\usepackage{color}

\newcommand{\om}{\omega}
\newcommand{\mmax}{\operatorname{max}}

\newcommand{\chisq}{\chi^2}

\newcommand{\chisqave}{\chisq_\text{ave}}
\newcommand{\chisqmin}{\chisq_\text{min}}

\usepackage{amssymb}

\newcounter{bla}

\journal{Computer Physics Communications}

\begin{document}

\begin{frontmatter}



\title{SAIL: A CUDA-based implementation of the simulated annealing for the inverse Laplace transform problem}


\author[a]{Yaroslav Lutsyshyn\corref{author}}
\author[b]{Grigori E. Astrakharchik}

\cortext[author] {Corresponding author.\\\textit{E-mail address:} yaroslav.lutsyshyn@uni-rostock.de}
\address[a]{Institut f\"ur Physik, Universit\"at Rostock, 18051 Rostock, Germany}
\address[b]{Departament de F\'isica, Campus Nord B4-B5, Universitat Polit\`ecnica de Catalunya, E-08034 Barcelona, Spain}

\begin{abstract}
We developed a CUDA-based parallelization of the annealing method for the inverse Laplace transform problem. 
The algorithm is based on annealing algorithm and minimizes residue of the reconstruction of the spectral function. 
We introduce local updates which preserve first two sum rules and allow an efficient parallel CUDA implementation. 
Annealing is performed with the Monte Carlo method on a population of Markov walkers. 
We propose {\em imprinted branching} method to improve further the convergence of the anneal. 
The algorithm is tested on truncated double-peak Lorentzian spectrum with examples of how the error in the input data affects the reconstruction. 
\end{abstract}

\begin{keyword}
Laplace transform, analytic continuation, Monte Carlo, CUDA, GPU.
\end{keyword}

\end{frontmatter}



{\bf PROGRAM SUMMARY}

\begin{small}
\noindent
{\em Program Title:} SAIL
(Simulated Annealing for the Inverse Laplace transform) \\
{\em Licensing provisions:} MIT\\
{\em Programming language:} CUDA-C                                  \\


{\em Nature of problem:} \\
Inverse Laplace transform is carried by means of repeated simulated annealing of the correlation data with a simple spectral model. Use of the GPU computing
allows for significant improvement in the computational throughput. 
\end{small}
\section{Introduction \label{sec:Introduction}}


The problem of the inverse transform arises in many different areas of sciences, including mathematics, physics, signal processing, image recognition, medical scanning, and many others \cite{RabinerGold1975}.
Even some animals (bats and dolphins) effectively rely on inverse transformation when they use echolocation for determining the position of objects in space
by using a special tool known as the sonar system \cite{Harrison738,Altes912}.
It is interesting that the processing of this information is so complicated, that in animal it is done by a dedicated part of its brain.
In the following, we aim to benefit from the computational power of GPU for solving the problem of Laplace inversion.

To be specific, we consider the spectral analysis formulation of the inverse problem, when the function of interest $S(\omega)$ is integrated over a known kernel $K(\omega,\tau)$ and the resulting function $F(\tau)$ is assumed to have been obtained from the experiments or numerical simulations,
\begin{align}
F(\tau) = \frac{1}{2\pi}\int_0^{\infty} S(\omega) K(\omega, \tau) d\omega.
\label{Eq:transform}
\end{align}
The problem of inversion consists in reconstructing the spectral function $S(\omega)$ given the knowledge of $F(\tau)$.
For some kernels $K(\omega,\tau)$, the reconstruction can be done relatively easily. For example, for the Fourier transform with $K(\omega, \tau) = \cos(\omega \tau)$, the inversion is straightforwardly obtained by the inverse Fourier transform. On the other hand, in the case of the also physically relevant Laplace transform with its exponentially decaying kernel, $K(\omega, \tau) = \exp(-\omega \tau)$, $\omega>0$, the problem of inversion becomes {mathematically ill-posed}.
This means that for a given signal $F(\tau)$, it is possible to find multiple spectral reconstructions $S(\omega)$ such that the transformation~(\ref{Eq:transform}) is satisfied within the same level of numerical accuracy.

The problem of the inverse Laplace transform being ill-posed is gravely exacerbated when the task is being solved purely numerically. The presence of a grid for $F(\tau)$ or $S(\om)$, and the limit on the computer level of numerical accuracy are both sufficient to produce a continuum of equally ``satisfactory'' inverse transforms. Additionally, the input data usually comes from other numerical calculations and contains a significant amount of statistical noise.

One of the typical problems arising in the inverse Laplace transform is the appearance of the {sawtooth instability} \cite{Prokof'ev13}. The reconstructed weights $S(\omega)$ with adjacent frequencies $\omega$ wildly oscillate, while they nonetheless provide a numerically good reconstruction for $S(\tau)$.
One possible way out is to demand continuity of $F(\omega)$, but unfortunately, {\it a priory} it is not known if $F(\omega)$ is continuous everywhere. 
On the contrary, it is quite typical that the spectrum contains a narrow peak. There are also known examples with discontinuous spectrum (for example, one-dimensional ideal Fermi gas).

Approaches based on {\em stochastic reconstruction} use the linear property of Eq.~(\ref{Eq:transform}), that is, that the transform~(\ref{Eq:transform}) of a sum of functions is a sum of their transforms.
In other words, by generating a large number of inverse transforms of the same input data and averaging over the reconstructions with good values of the residue, typically it is possible to alleviate the problem caused by the sawtooth instability. Another good feature of the stochastic averaging is that it provides a measure on the errorbars of the reconstruction.

A popular approach which produces rather smooth results is the {\em maximal entropy} method \cite{Silver90,Jarrell96,Bryan90,Gubernatis91,Dirks10,Gunnarsson10,Gunnarsson10b}.
Within this technique one formulates a target class of functions and ensures that the found reconstruction is the best within this class. The function which is minimized consists of the residue with an additional ``entropy'' term, which is weighted by an external parameter, sometimes associated with an effective temperature. An intermediate value of this parameter ensures that both the quality of the reconstruction and the smoothness through likeness to a chosen model are simultaneously optimized.
Although the maximal entropy method is very fast, it has a drawback in that due to the imposed constraints, an important systematic bias may be introduced.
The reconstruction depends on the model and on the exact value of the external parameter (effective temperature). Additionally, it is not obvious to which extent the resulting errorbars reflect a possible bias due to a particular choice of the model.

Method of {\em consistent constraints} \cite{Mishchenko00,Mishchenko12,Prokof'ev13}
aims at finding a reconstruction which at the same time is locally smooth and has a rather good residue.
This method incorporates four penalties: optimizing residue, first derivative (avoids sawtooth instability), amplitudes (searches for a non-negative solution), 
and deviations from the target form (chosen to stabilize the second derivative).
The regularization constants in the penalties are adjusted consistently with the solution using a feedback protocol during the iterations.

Another recent approach dealing with the problem of inversion is based on the so-called genetic inversion via falsification of theories (GIFT) strategy \cite{Vitali10,Vitali09}.
Two main ingredients of this approach are the falsification principle and the genetic dynamics. A number of models are generated such that none of them falsify the requested properties, including the sum rules.
The averaging over the resulting models is used to remove the spurious differences while enhancing the physical information.
The ``genetic'' part of the method refers to the evolution-like processes (selection, crossover and mutation of reconstructions) that are used to evolve the subsequent generations.

The large number of different approaches dealing with the inverse Laplace problem underlines the importance of this task to the Physics community, and the fact that no universal consensus has been reached as to which method provides the most reliable result. In this work, we implement what is essentially a brute-force approach to the inverse transform. We employ a very general piecewise spectral model, and assume little about the properties of the spectrum beyond the first two sum rules. We thus aim to put some light on the extend to which the inverse Laplace transform can be carried on a computer without imposing any particular restraint on the nature of the resulting spectrum. This limit, one may hope, is changing as more powerful computing architecture becomes available. We use the method of simulated annealing to minimize the residue of the transform, and build the final spectrum from a large number of transforms computed in parallel on a graphical processing card (GPU). The program is written in CUDA as is made available to public. Interested readers are welcome to use this program with limits to the spectral model that are specific to their particular systems of interest.

\section{Simulated annealing}

\subsection{Spectral model}

It is assumed that the correlation data $\tilde{F}(\tau)$ is available on a uniform grid of size $N_\tau$,
\begin{equation}
\tau_i =  i\Delta\tau, \quad i=0,\dots,N_\tau-1.
\end{equation}
In the case of dynamic form factor, $S(k,\tau)$, each $k$-vector can be transformed independently, thus for simplicity we will omit the $k$ index in the following.
The \emph{spectral model} in the frequency domain is taken to be a piecewise-rectangular function,
\begin{equation}
S(\omega)=\sum_{j=0}^{j=N_\omega-1} a_j \Theta(\omega-j\Delta\omega)\Theta(j\Delta\omega+\Delta\omega-\omega).
\label{eq:spectral-model}
\end{equation}
We start the indexing from zero to be consistent with the  computer program.
The real-valued nonnegative coefficients $a_j$ define the spectral model in the range $\omega\in[0,\Delta\omega N_\omega]$.
The \emph{correlation model} in the imaginary-time domain is therefore given by
\begin{align}
F(\tau) &=  \frac{1}{2\pi}\int_0^{\om_{\mmax}} S(\om) \ee{-\om \tau} d\om  \\
        &= \frac{1}{2\pi}\sum_{j=0}^{N_\om-1} a_j
                  \int_{j\Delta\om}^{(j+1)\Delta\om}\ee{-\om\tau}d\om     \\
        &=  \frac{1}{2\pi} \frac{1-\ee{-\Delta\om\tau}}{\tau}
                          \sum_{j=0}^{N_\om-1} a_j \ee{-j\Delta\om\tau}.
\label{eq:correlation-model}
\end{align}
Terminating the above integral at $\om_{\mmax}=\Delta\omega N_\omega$ enforces the assumption that the spectral model $S(\om)$ is equal to zero everywhere above the maximal frequency.
Although in general the spectrum might have an infinite support, physically the containing energy is finite and is mostly contained in a finite range of frequencies.
In practice, the assumption of the maximal frequency works well and always can be checked by making the ceiling higher.
The evaluation of correlation model~(\ref{eq:correlation-model}) is numerically efficient. One can write
\begin{equation}
F(\tau_i) = c_i \sum_{j=0}^{N_\om-1} a_j \left(\ee{-\Delta\om\Delta\tau}\right)^{ij},
\label{eq:efficient-correlation-model}
\end{equation}
with
\begin{equation}
c_i=
\begin{cases}
\frac{\Delta\om}{2\pi}, & i=0 \\
\frac{1-\ee{-i \Delta\om\Delta\tau}}{2\pi i \Delta\tau}, & i>0.
\end{cases}
\label{eq:c-coefficients}
\end{equation}
We define the residue, or error, function as
\begin{equation}
\chisq = \sum_{i=0}^{i=N_\tau-1}  \left( F(\tau_i)-\tilde{F}(\tau_i) \right)^2.
\label{eq:residue}
\end{equation}
The residue is a nonnegative function defined in the space of coefficients $a_j$. Given the large size of this space, it is troublesome to minimize $\chisq$ directly. Instead, we follow the prescription of the simulated annealing method~\cite{LandauBinderBook}.
We interpret the function
\begin{equation}
P=\exp\left[-\chisq(a_0,\dots,a_{N_\om-1})/T\right]
\prod_j\Theta(a_j)
\label{eq:probability-density}
\end{equation}
as a multivariate probability density for the spectral coefficients $a_j$ with $T$ being the effective temperature. In the limit when the parameter $T$ is small, the likely values of the coefficients are limited to those which minimize $\chisq$. The assumption of the annealing method is that one may begin to sample probability density~(\ref{eq:probability-density}) with large $T$, where a broad range of  values of $a_j$ is allowed by the probability density, then to gradually decrease $T$ to compress the allowed space of $a_j$ close to the region where one minimizes the residue function.
That is, we aim to compute the ``frozen'' limits $a^\ast$, defined as
\begin{equation}
a^\ast_j  = \lim_{T\rightarrow 0} \frac{\int a_j \exp(-{\chisq}/{T})da_0 \cdots da_{N_\om-1} } {\int \exp(-{\chisq}/{T})da_0 \cdots da_{N_\om-1} }.
\label{eq:residue-average}
\end{equation}
Fittingly for multidimensional integration, the above expression is evaluated with the Monte Carlo method. By analogy with the classical statistical mechanics, we refer to $T$
as the annealing temperature, although in our case its meaning is purely mathematical. The manner in which one lowers the annealing temperature is called the annealing prescription. We will describe our prescription further below.

\subsection{Update scheme}

The annealing method relies on an efficient sampling of the probability density given by Eq.~(\ref{eq:probability-density}). We carry this sampling with the Metropolis method~\cite{Metropolis-1949}. The progress of the Markov chain benefits from an efficient choice of updates. We have designed the moves to be simultaneously numerically efficient and to preserve the first two sum rules. For our spectral model, the sum rules are satisfied if the moves preserve two sums on the spectral coefficients $a_j$,
\begin{equation}
\int S(k,\om) d\om = \Delta\om \sum_j a_j = \Delta\om A,
\end{equation}
and
\begin{align}
\int S(k,\om) \om d\om
 & = \sum_j a_j \frac{(\om_j+\Delta\om)^2-\om_j^2}{2} \\
 & = \Delta\om \sum_j a_j (\om_j+\Delta\om/2) \\
 & = \Delta\om B,
\end{align}
where we have defined the sum constants $A$ and $B$. The first constant is known directly from the input as one can notice from Eq.~(\ref{eq:efficient-correlation-model}),
\begin{equation}
\tilde{F}(0)=c_0 A.
\end{equation}
Constant $B$ requires the knowledge of the $k$-vector at which the correlation data was collected and thus must be provided as an additional input to the program. We intent to perform the Metropolis updates in the restricted space where both $A$ and $B$ sums are constant. To this end, we select two indices, $p$ and  $s$, and two amplitudes, $\delta_p$ and $\delta_s$. The spectral model is updated according to
\begin{align}
a_p' &= a_p r + \delta_p \label{eq:update-scheme-p} \\
a_s' &= a_s r + \delta_s \label{eq:update-scheme-s}\\
a_j' &= a_j r, \quad j\ne s,p. \label{eq:update-scheme-j}
\end{align}
It is straight-forward to verify that both sum rules are preserved if one satisfies
\begin{align}
\delta_p (\om_p-\om_s) &= [B-A(\om_s+\Delta\om/2)] (1-r) \label{eq:update-delta-p}\\
\delta_s (\om_s-\om_p) &= [B-A(\om_p+\Delta\om/2)] (1-r) \label{eq:update-delta-s}.
\end{align}
We select the value of $r$ as our ``random move'', which determines the values of $\delta_p$ and $\delta_s$.

The moves that would result in a negative spectral model are rejected according to the definition of the probability density given in Eq.~(\ref{eq:probability-density}). The strength of the update scheme~(\ref{eq:update-scheme-p})--(\ref{eq:update-scheme-j}) becomes apparent when one notices that the updated correlation model is given by
\begin{equation}
F'(\tau) = r F(\tau) + \delta_p\ee{-\omega_p\tau}+\delta_s\ee{-\omega_s\tau}.
\end{equation}
Thus each update requires only $O(N_\tau)$ calculations and even does not require memory loads for the spectral coefficients $a_j$. To update the entire model, one needs $O(N_\om)$ single updates. Thus a macroupdate requires only $O(N_\tau N_\om)$ calculations.
The ergodicity of the update scheme can be easily verified for the case of small update amplitudes, which is always the case towards the end of the annealing procedure.

At first glance, this update scheme introduces a special sampling frequency,
\begin{equation}
\om_c=B/A.
\label{eq:special-frequency}
\end{equation}
According to Eqs.~(\ref{eq:update-delta-p}) and (\ref{eq:update-delta-s}), if one of the updated frequencies is close to $\om_c$, amplitude update to the spectral density at the other frequency will be suppressed. However, the amplitude update at the $\om_c$ itself is not suppressed. It is the update at the other frequency that would be suppressed on such an occasion. However, the pairs of $\om_p$ and $\om_s$ are selected independently of each other and thus the effect, if any, is spread throughout the spectrum. 
Whenever we do find any artifacts around $\om_c$, they are usually on the same level of magnitude as other artifacts in the system.

\subsection{Selection of update frequencies}

We have found that our annealing scheme is more efficient when we update more frequently the elements that contain larger spectral density. On the other hand, frequencies with small spectral coefficients $a_j$ must also be updated, although perhaps less often.  The spectral coefficients are nonnegative and can be interpreted as probability density. For each move, we perform a truncated Markov walk in the space of the spectral coefficients, and thus select the indices $p$ and $s$. If the walk has zero length, indices $p$ and $s$ are selected randomly and uniformly on $[0,N_\om-1]$. In the limit of infinitely long walks, the indices are selected according to probability $P(j)=a_{j}/A$. The truncated walk provides a convenient mixture of the two distributions. Additionally, a small fraction of the uniform distribution can be mixed in.

\subsection{Self-cooling \label{sec:self-cooling}}

We found it rather difficult to provide a universal cooling prescription for the different correlation data that we have had considered. Instead, we developed a ``self-cooling'' prescription in which the temperature is lowered according to the residue to which the system has already been able to relax. Suppose $\chisqave$ is the average residue of the annealing population, and $\chisqmin$ is the smallest one. In the beginning, we set $T$ close $\chisqave$. In this regime, the system ``boils'' to the largest residues allowed under the restraint of the sum rules and the limited number of the degrees of freedom. After a certain number of steps, the temperature is slowly ramped towards $\chisqmin$. This is usually accompanied by a sharp decrease in the residue. Such a cooling stage is followed by a freeze-out as the temperature is lowered below the minimal residue in the ensemble. The freeze-out has temperature lowered proportionately to the the number of steps carried.

\subsection{Imprinted branching and quenching the sawtooth instability \label{sec:branching}}

The annealing is carried on an ensemble of $N_w$ walkers, each representing a separate Markov chain of the Metropolis Monte Carlo described in the above subsections. We have found that the annealing process is improved if we incorporate branching between the walkers, similar to the branching of walkers in the Green's Function Monte Carlo \cite{Wells1987-GreensFunctionMonteCarloMethods} methods. This idea was inspired by the replication of information in the GIFT method \cite{Vitali10,Vitali09}, and by the recent success of the simulated quantum annealing methods. Branching made a considerable difference in minimizing the residue of the resulting transform. However, several attempts to map the system to a quantum Hamiltonian and carry a full simulated quantum annealing did not lead to an improvement. For this reason, we carry an annealing process in which sampling is carried with the Metropolis method as described above, i.e. sampling the probability density given by Eq.~(\ref{eq:probability-density}). After a certain number of Metropolis steps, weights are assigned according to Eq.~(\ref{eq:probability-density}),
\begin{equation}
w=\exp[-(P(\chisq)-P_r)\Delta],
\end{equation}
with two constants $P_r$ and $\Delta$ as explained below.
Each walker is assigned an integer multiplier $m$ according to
\begin{equation}
m=\left\lfloor{w+\xi}\right\rfloor,
\end{equation}
where $\xi$ is a random number uniformly distributed on the unit interval and $\lfloor{\cdots}\rfloor$ denotes the floor function. The expectation value of the multiplier $m$ equals to the weight $w$. 
For each branching event, the parameter $P_r$ is adjusted in such a way that, 
for a once generated set of random numbers $\xi$,
the sum of all multipliers equals to the number of walkers, i.e.~fixing the population of walkers. Parameter $\Delta$ is found numerically such that a given fraction of walkers is assigned a zero multiplier, $w=0$.
We refer to this fraction of walkers as the \emph{branching fraction}.
These walkers are to be replaced by the walkers with $m>1$.

Linearity of the transform (and the sum rules) allows for a unique branching in which the walkers are not completely removed and replaced by ``better'' walkers but instead are transformed to a linear combination of their own spectral density and the density from the replacing walker. We refer to this process as {\em imprinting}.
Let the index of one of the walkers with zero multiplier be $l_1$ and its spectral model $\{a_j^{(l_1)}\}$; let $l_2$ be the index of the walker that was selected to imprint on $l_1$. The result of imprinting is a change in the spectrum of $l_1$,
\begin{equation}
{a_j^{(l_1)}\leftarrow t {a_j^{(l_2)}} + (1-t) {a_j^{(l_1)}}  },
\end{equation}
and no change in the spectrum of $l_2$. The imprint strength $t$ can be set constant or adjusted during the execution. In the limit $t=1$ one recovers traditional branching.

As the system freezes to a minimum of the residue, each walker begins to develop sawtooth instability. Branching can then propagate a particular instability through the entire population of walkers. An example of this effect is shown further below. We find that it is possible to counter most of this effect by reducing the imprinting strength and the branching fraction according to the decrease of the acceptance probability. With less branching, the population effectively self-averages as random moves are taken around the global minimum. The average is then enforced on the population through the imprinting.

\subsection{Smoothing \label{sec:smoothing}}

The adaptive branching strength described in the above Section significantly delays the onset of the sawtooth instability, but does not prevent its eventual formation. To counter this effect, we introduced a three-point stencil for smoothing the transform, according to
\begin{align}
a_j&=sa_{j-1}+(1-2s)a_j+sa_{j+1},\quad 0<j<N_\om-1 \\
a_{0}&=(1-s)a_0+sa_1  \\
a_{N_\om-1}&=(1-s)a_{N_\om-1}+sa_{N_\om-2} ,
\end{align}
which we carry with a rather small smoothing amplitude $s$, usually of the order of several percent. The smoothing stencil preserves the first sum rule, but requires an additional renormalization due to a violation in the second sum rule of the order of $s (a_{N_\om-1}-a_0)/N_\om$, which is quite small both because the spectral density on the endpoints tends to vanish, and because $s/N_\om \ll1$.

It is important to point out that smoothing here is not a form of post-processing, but is instead a part of the algorithm itself. Once the transforms are smoothed and renormalized, they undergo Metropolis updates and branching. The process is then repeated. Walkers in which smoothing were to produce unfavorable residue values are branched away.

\subsection{Update amplitude heuristics}

As the residue function approaches its minimum, and the annealing temperature is significantly decreased, 
the acceptance rate of the proposed moves decreases.
To improve sampling at this stage, one may decrease the width of the distribution of the displacement of $1-r$. We choose to scale the displacement amplitude by an additional factor $y$ which is computed according to the ongoing acceptance rate $a$ as
\begin{equation}
y=
\begin{cases}
1, & a\ge p \\
(a/p)^m , & a<p,
\end{cases}
\label{eq:amplitude-acceptance-adjustment}
\end{equation}
where one selects parameters $m$ and $p$, $m>1$ and  $0\le p\le 1$. The choice of parameters is discussed below.

\section{Choice of parameters}

This inversion code aims to require minimal user and problem-specific input, even at expense of additional computational effort. This requires a number of choices to be fixed ``out-of-the-box''. However, if required, most of the parameters discussed in this Section can be readily adjusted.

The number of nodes in the correlation data $N_\tau$ and the timestep $\Delta\tau$ are fixed by the user in the input data.
The user must also select the spectral domain, by specifying $\om_\text{max}$ and $N_\om$.

The ``natural'' scale for the annealing temperature $T$ is determined by the value of the residual, which is in turn determined by the input data.
Before the annealing, the program computes the starting residue value. It is then used as a unit of both annealing temperature and the residue.
Thus the starting (scaled) residue is always equal to unity.
We choose the starting annealing temperature as equal to $T=10$, which results in the initial boiling of the system.

The annealing process in organized into blocks. Each block consists of a number of Metropolis steps. To update the entire spectral model, one should carry about $N_\om$ such steps. Each block thus consists of some $kN_\om$ steps. We select $k=100$.

Each block is followed by the branching step. We select to branch a quarter of all walkers on the branching step, combined with a small imprinting strength equal to $t=0.1$. 

After some experimentation, we found that the parameter $p$ from Eq.~(\ref{eq:amplitude-acceptance-adjustment}) might as well be set to $p=1$. We also select $m=2$.

\begin{figure}[t]
\begin{centering}
\includegraphics[width=\textwidth]
{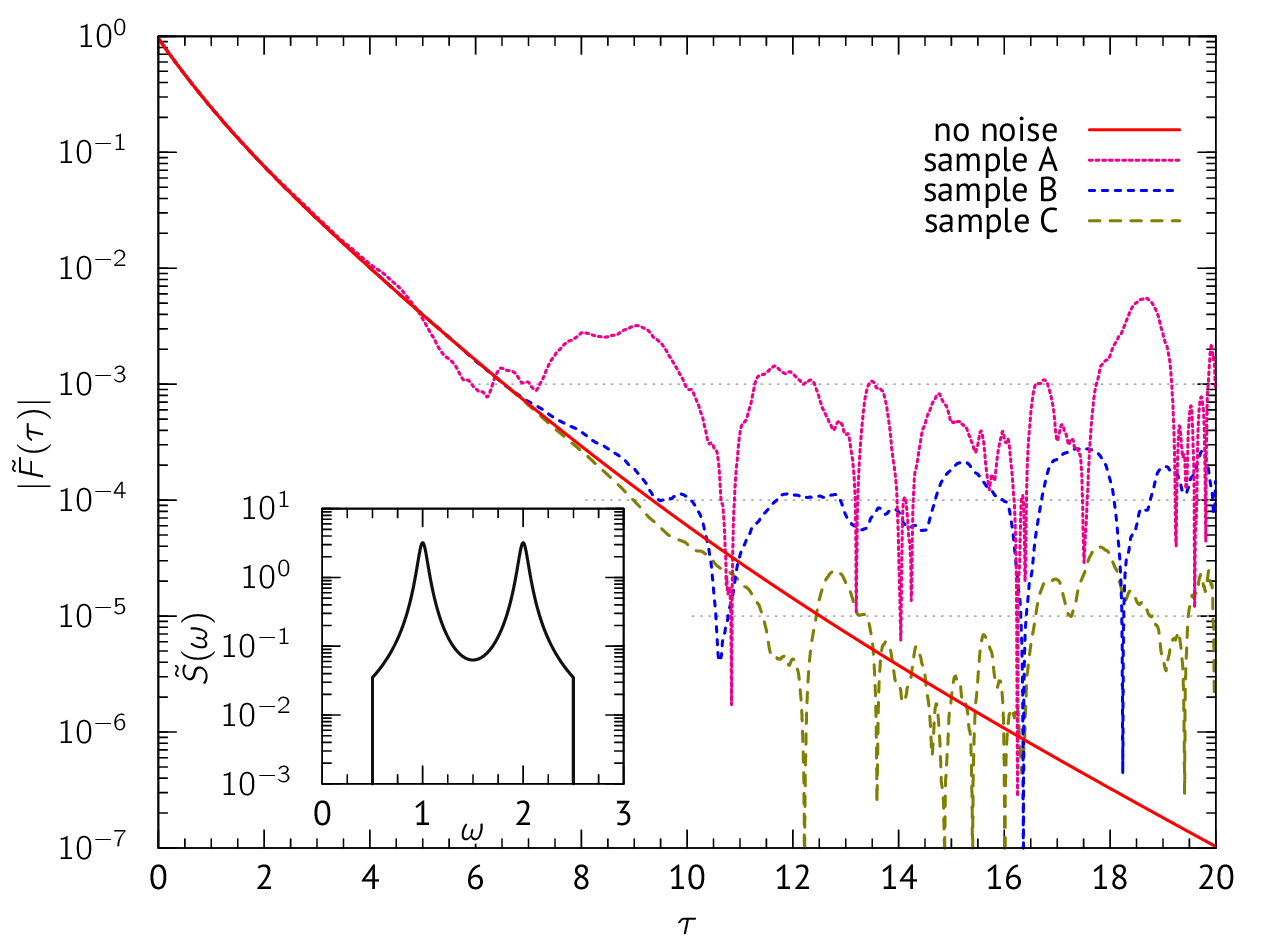}
\caption{ Examples of generated input correlation data for the Lorentzian band model of Eq.~(\ref{eq:lorentzian-band}).
Solid line shows the numerically precise correlation data.
Samples A through C show data resampled as outlined
in Section~\ref{sec:resampling}.
These samples correspond to the r.m.s.\ noise level
close to, correspondingly, $10^{-3}$,  $10^{-4}$, and  $10^{-5}$.
The lowest noise level required sampling of $3\cdot10^{8}$
correlated sequences.
The inset shows the spectral density that was used to build
the correlation data.
\label{fig:lorentzian-noise}
}
\end{centering}
\end{figure}

\section{Sample execution}

The package has been maximally streamlined and simplified for the distribution. 
It includes just three files, \texttt{sail.cpp} which includes all of the host functions and GPU kernels, \texttt{sail.h} which is simultaneously a header file and the file which contains certain execution parameters, and \texttt{main.cpp} file which shows how to invoke the transform. All the relevant methods are wrapped in a single C++ class called \texttt{ILaplace}. The user is expected to invoke a single instance of this class for a given transform, call parameter setters and finally invoke the anneal method. All of this is demonstrated in \texttt{main.cpp}. A makefile is included to aid compilation. Since most of the parameters are set through the preprocessor macros, it is expected that recompilation will be often necessary.

Upon unpacking, the code may be compiled on a CUDA-enabled machine by invoking
\begin{verbatim}
make sail
\end{verbatim}
The program requires Nvidia GPU with compute capability of at least 3.5 (i.e.\ Kepler generation) or larger. 
The directory contains a file named \texttt{corr.in} which is the data for sample C from the next section. The annealing may be invoked by executing
\begin{verbatim}
./sail 
\end{verbatim}
As the annealing proceeds, the code will print current values of temperature, residue, and the acceptance rate. A sample output is located in a subdirectory, in file \texttt{sample-run/log-1500}. The sample was obtained on a K40 card. The same subdirectory contains other input data that was used in Section~\ref{sec:intro-toy}.

This sample execution will store the ongoing spectrum after each execution block, to the file named \texttt{spectrum.dat} and to a file named \texttt{spectrum-\%5d.dat}, where the five-digits number space \texttt{\%5d} contains the block number.
Once the program reaches the anneal minimum, it will terminate and store the final spectrum into file \texttt{spectrum-fin.dat}.
A number of example spectra are placed in directory \texttt{sample-run}.

Users are encouraged to explore the header file, \texttt{sail.h}, to see the adjustable parameters. In particular, if the user's GPU lacks in memory capacity, it may be necessary to reduce the number of walkers defined in the macro \texttt{Ninstance}.

\section{Annealing example\label{sec:intro-toy}}

\subsection{Correlation data generation \label{sec:resampling}}

Here we present the results of the inversion obtained for a toy model. 
The input is exact to double numerical precision and thus by far exceeds any input that may be possible with real-world data. 
However, the number of grid points and the extend of the time domain are both limited. 
Thus there is a continuum of spectral models that can fit this ``exact'' correlation data. 
Studying such a case allows one to understand the fundamental limitations of the particular inversion algorithm, before they are masked by inaccuracy of the input data.

We also invert toy models which simulate noisy input data. The outcome of typical Monte Carlo simulation for the correlation data contains strongly correlated statistical noise. Such noise arises primarily from undersampling of the density-density correlation. The errors that originate from the algorithm itself are instead systematic in nature. Characteristically, undersampling is also more severe for larger values of $\tau$. We emulate this effect as follows. Starting from the known model $\tilde{S}(\om)$, we compute the multivariate Gaussian distribution which has the autocorrelation equal to the one we desire. From this distribution we draw a limited number of instances, where each instance is itself a correlated Gaussian sequence. We compute the autocorrelation of this limited set, which thus has the undersampling noise built into it. This process results in correlation data with a realistic noise profile. Noise level can be varied by adjusting the number of sampled sequences.
An example of resulting correlation data is shown in Fig.~\ref{fig:lorentzian-noise}.

\subsection{Two-peak Lorentzian band \label{sec:lorentzian-model}}

For our model, we select a double-peak Lorentzian spectrum that is additionally cut to a strip of frequencies, that is
\begin{equation}
\tilde{S}(\om) = [L(\om;\om_1,\Gamma_1)+L(\om;\om_2,\Gamma_2)]
\Theta(\om-\om_{l})
\Theta(\om_{h}-\om),
\label{eq:lorentzian-band}
\end{equation}
where $L(\om;\om_0,\Gamma)$ is a Lorentzian centered around $\om_0$ with half-width at half-maximum given by $\Gamma$.
We use Lorentzians of equal width, $\Gamma=1$, and located at $\om_1=1$ and $\om_2=2$. Band is clipped at $\om_l=0.5$ and $\om_h=2.5$. The resulting function  can be seen in the inset to Fig.~\ref{fig:lorentzian-noise}.
Such a structure allows to study several aspects of the inversion algorithm. The exact shape of the second peak is known to be difficult to reconstruct \cite{SvistunovTBP}. The low but non-vanishing region between the peaks allows to see the ``dynamic range'' of the method, that is, how well does the method resolve the low amplitudes that are present in conjunction with the large ones. Finally, the band clippings at $\om_l$ and $\om_h$ provide a well-defined frequency domain but are challenging to resolve as they occur at low amplitudes.

\subsection{Annealing results}

We carried the annealing procedure following the described above protocol. The frequency domain was limited to $\om<3.5$ and the spectral model was intentionally oversampled. We use one thousand points in the frequency domain. In this regime, there is little observable dependence on the number of frequency nodes. All the results presented here were obtained with $1500$ walkers.

\begin{figure}[ht]
\begin{centering}
\includegraphics[width=\textwidth]
{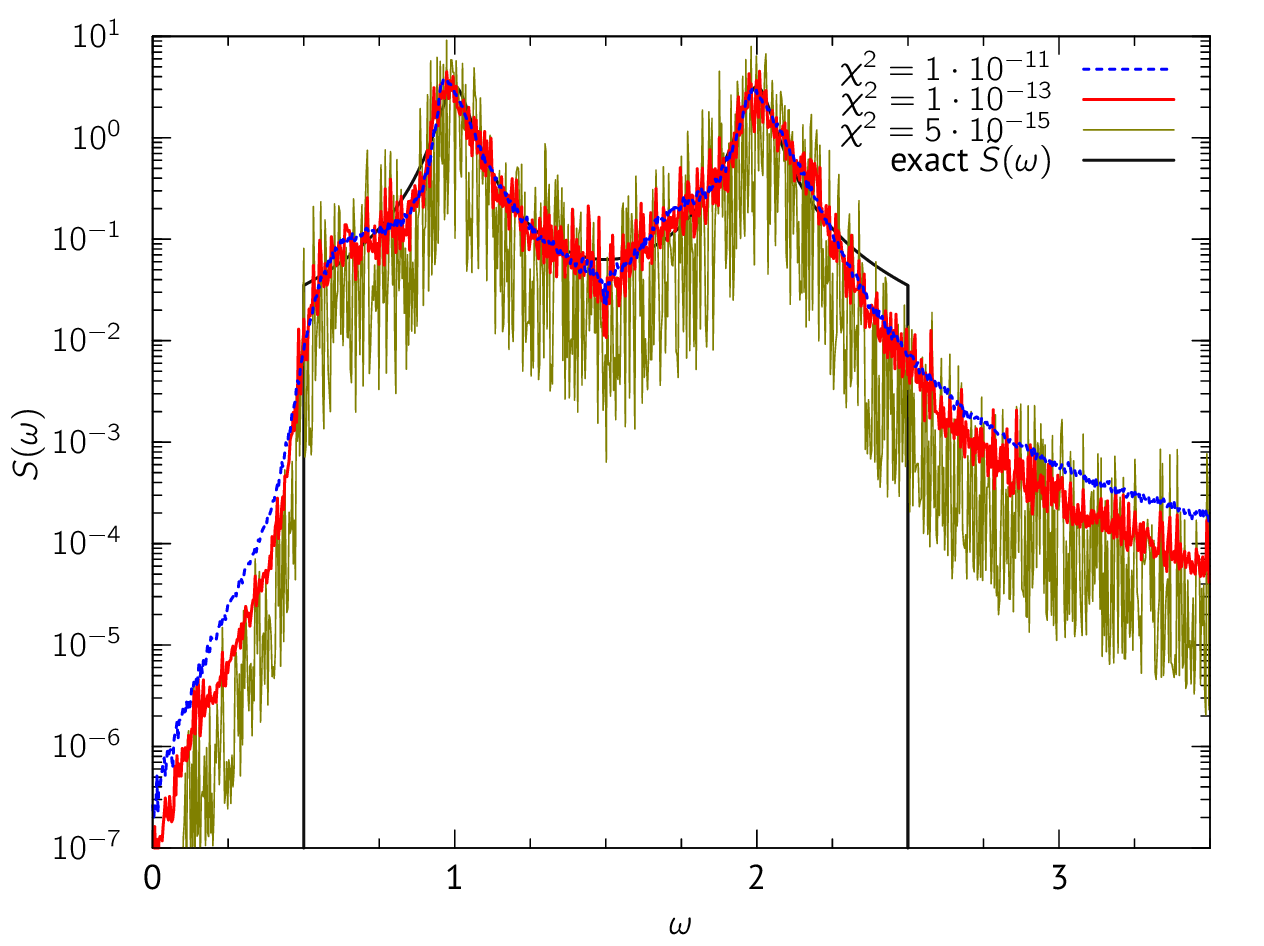}
\caption{
Example of sawtooth instability. The reconstruction was carried on noiseless Lorentzian band as described in Section~\ref{sec:lorentzian-model}. The reconstruction used imprinted branching as described in Section~\ref{sec:branching}, but employed neither adaptive branching nor smoothing. The transforms shown here are from the same sequence, marked by their decreasing value of the residue function $\chisq$. Once the transform reached the residue value of $10^{-11}$, it has mostly converged. However, further ``improvement'' followed with the residue dropping to below $10^{-14}$. This gain came mostly from the sawtooth pattern which developed in the spectrum. That is, such a pattern fits the grid input data better.
\label{fig:sawtooth}
}
\end{centering}
\end{figure}

\begin{figure}[ht]
\begin{centering}
\includegraphics[width=\textwidth]
{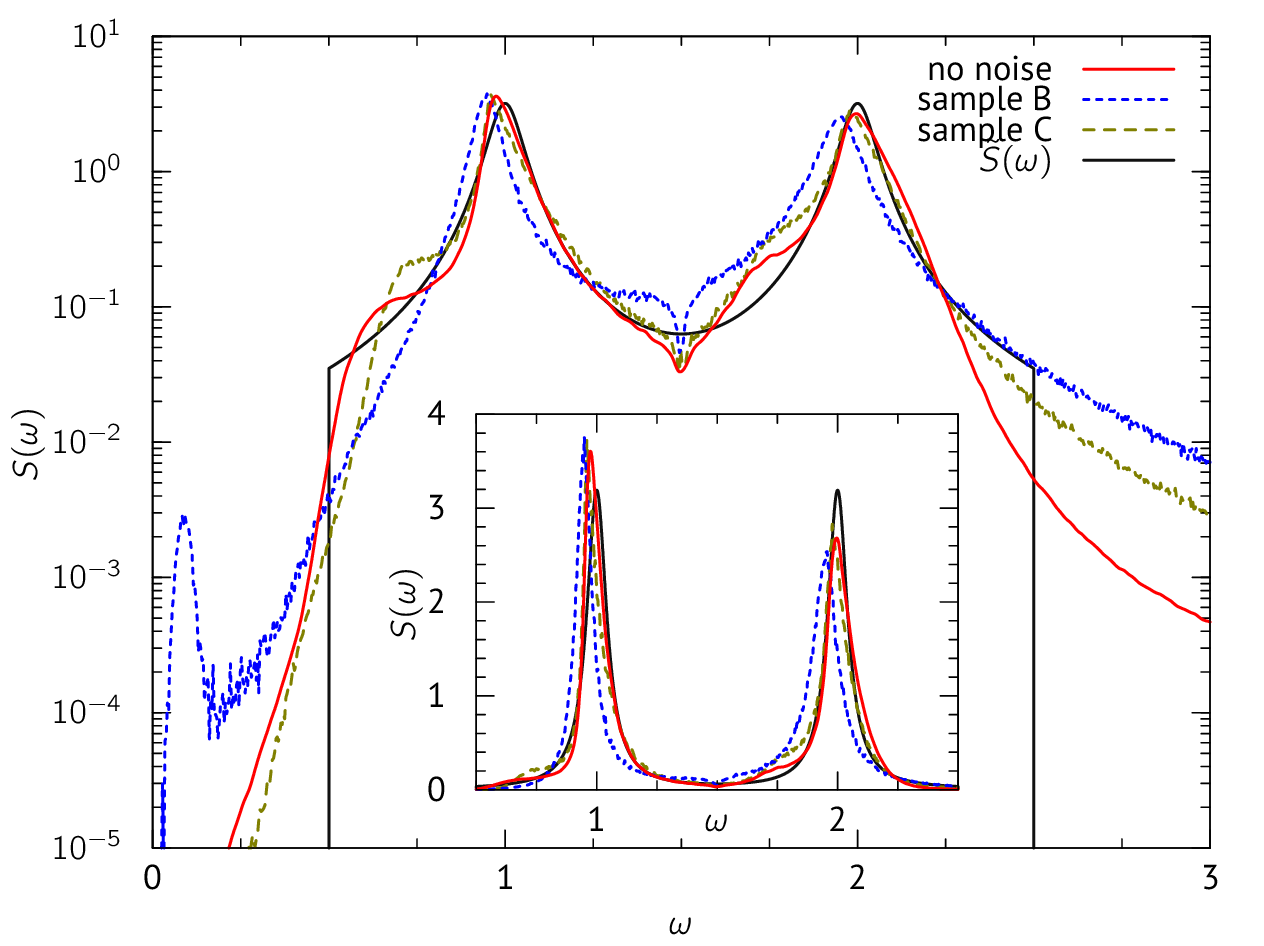}
\caption{
Anneal transforms for different noise levels. The input data is shown on Fig.~{\ref{fig:lorentzian-noise}}. One can notice deterioration of the transform quality. The low-level features, such as the lower cusp frequency, are not reproduced. The shape of the peak is increasingly deteriorated. Furthermore, sample B dataset, with noise at the level of about $10^{-4}$, developed a phantom low-frequency feature, albeit with low intensity. 
Transform for sample A, with noise level $10^{-2}$, failed to discern the second peak and is not shown here.
The inset shows the transforms on the linear scale.
\label{fig:anneals}
}
\end{centering}
\end{figure}

Annealing with fixed branching ratio and imprint strength $t=0.1$, as described in Section~\ref{sec:branching}, revealed the sawtooth instability. The corresponding transforms are shown in Fig.~\ref{fig:sawtooth}. It is worthwhile to notice that the transformation has mostly converged before the instability started to develop.
Thus in some cases, user may prefer to terminate the transform early.
To avoid the instability, below we have used both the adaptive branching and smoothing as described in Sections~\ref{sec:branching} and~\ref{sec:smoothing}. The smoothing parameter was selected to be $10^{-2}$. Because of its small amplitude, and the large number of steps taken in each block ($100N_\om=10^5$), the effect of smoothing did not appear until the late stages of annealing.

We further tested the method with the artificially noisy input data, built as described above and shown in Fig.~\ref{fig:lorentzian-noise}. The resulting transforms are shown in Fig.~\ref{fig:anneals}. We find that the for the noise levels under $10^{-2}$ (samples B and C), both peaks are discovered by the method. The location and shape of the peaks is increasingly distorted with increasing noise, until the peaks are no longer discernible (transform for sample A with noise level $10^{-2}$ is not shown in Fig.~\ref{fig:anneals}). Position of each peak is determined with certain error, which increased with the increasing noise level. For Sample B data, error in the peak location reached 0.05, i.e.~as much as $5\%$ for the first peak. The heights of the peaks are not reproduced very well, which is quite visible on the linear scale (see the inset in Fig.~\ref{fig:anneals}). We compared the discrepancy in strength of the peaks. Here we define the skew as
\begin{equation}
x=\frac{\left| \int_0^{\om_m} S(\om)d\om
              -\int_{\om_m}^\infty S(\om)d\om \right|}
{
\int_0^\infty S(\om)d\om
},
\end{equation}
where $\om_m=(\om_l+\om_h)/2=1.5$ is the midpoint between the peaks in Eq.~(\ref{eq:lorentzian-band}). 
We found that for the noiseless data, the skew is rather small, $x=0.004$. For noise level $10^{-5}$, we found $x=0.017$ and for noise level of $10^{-4}$, $x=0.05$. That is, the absolute strength of the peak is a property that is reproduced much better than its shape, a property of the inverse transform reported and studied in detail in \cite{SvistunovTBP}. Indeed, it is possible to find properties that map from ill-posed transforms but are themselves
well-posed~\cite{Franklin1970}.
Finally, it is worth noting that the upper cusp of the band $\om_h$ was not detected by either transform, while the lower cusp $\om_l$ was only detected by the transform of the noiseless data. In all transforms, it is possible to notice the feature around the special sampling frequency, which in our case lies exactly in-between the peaks. However, the distortion introduced at this frequency is in fact of the same magnitude as the distortion introduced around the lower cusp edge etc.


\section{Conclusions}

To conclude, we addressed the problem of the inverse Laplace transform. 
The main achievements are twofold:
(i) we have proposed the {\em imprinted branching} strategy to improve the quality of the reconstructed spectral function;
(ii) we have implemented an efficient CUDA-based parallel code that is being released as an open-source software.

In order to demonstrate the efficiency of our code, we have considered a complicated example of a truncated double-peak Lorentzian spectrum. 
We show how the noise in the input data affects the reconstruction. 
For the noise $10^{-2}$ the second peak is absent, while with $10^{-3}$ and $10^{-4}$ the reconstruction is getting progressively better. 
The inversion scheme is based on the annealing technique where the effective temperature is introduced in order to deal efficiently with the minimization of the $\chisq$ function which has a large number of degrees of freedom.
To improve the accuracy further we propose the method of imprinted branching in which the spectrum is replicated into an ensemble of {\em walkers}.
According to the goodness of the reconstruction of the walkers, each of them may produce different number of descendants including one (the walker is simply copied), zero (the walker is killed), two and more (i.e. it gets replicated). 
In the latter case, the branching is done using a mixture of walkers, so that some properties of one walker is imprinted on the other. 
When very low values of the $\chisq$ are reached, the sawtooth instability appears. 
In order to deal with it we use a smoothing method.

We have developed and tested a CUDA program for an efficient inverse Laplace transform. 
We create such an algorithm that all updates are local using to its best the CUDA architecture. 
Furthermore, we write the updates in such a form that the first two sum rules are preserved, incorporating the known physical information. 
Importantly, we do not impose any restriction on the spectrum so that the code has a very general applicability. 
The use of CUDA is extremely well suited to this heavy-number crunching problem, resulting in a community-friendly code. 
A computational power of s single GPU card may be comparable to that of several modern multicore CPUs, an is equivalent to having a small cluster in a single PC.
We hope that our package will be useful for scientists who have to deal with the inverse Laplace transform.

\section*{Acknowledgments}

Authors thankfully acknowledge the computer resources and assistance provided by the Spanish Supercomputing Network (Red Espa\~nola de Supercomputaci\'on). We thank the Nvidia Corp. for provided hardware and support. 
G.E.A. acknowledges support by the MICINN (Spain) Grant No. FIS2014-56257-C2-1-P.











\end{document}